# Photoplethysmography based atrial fibrillation detection:

# an updated review from July 2019

Cheng Ding*[1,2], Ran Xiao*[1], Weijia Wang[1], Elizabeth Holdsworth[3], Xiao Hu[1,2,4]

[1]Nell Hodgson Woodruff School of Nursing, Emory University, Atlanta, GA, USA

[2]The Wallace H. Coulter Department of Biomedical Engineering, Georgia Institute of Technology, Atlanta, GA, USA

[3]Georgia Tech Library, Georgia Institute of Technology, Atlanta, GA, USA

[4]Department of Biomedical Informatics, Emory University School of Medicine, Atlanta, GA, USA

*These authors contribute equally to this work and share the first authorship.

## Abstract

Atrial fibrillation (AF) is a prevalent cardiac arrhythmia associated with significant health ramifications, including an elevated susceptibility to ischemic stroke, heart disease, and heightened mortality. Photoplethysmography (PPG) has emerged as a promising technology for continuous AF monitoring for its cost-effectiveness and widespread integration into wearable devices. Our team previously conducted an exhaustive review on PPG-based AF detection before June 2019. However, since then, more advanced technologies have emerged in this field. This paper offers a comprehensive review of the latest advancements in PPG-based AF detection, utilizing digital health and artificial intelligence (AI) solutions, within the timeframe spanning from July 2019 to December 2022. Through extensive exploration of scientific databases, we have identified 59 pertinent studies. Our comprehensive review encompasses an in-depth assessment of the statistical methodologies, traditional machine learning techniques, and deep learning approaches employed in these studies. In addition, we address the challenges encountered in the domain of PPG-based AF detection. Furthermore, we maintain a dedicated website to curate the latest research in this area, with regular updates on a regular basis.

1. Introduction

AF is a highly prevalent cardiac arrhythmia, which affects approximately 1-2% of the general population, and is expected to continue to rise in the future worldwide due to population aging [1-3] . Individuals with AF face a substantially heightened risk of experiencing cerebral and cardiovascular complications. Specifically, they are at a fivefold higher risk [4] of ischemic stroke and are associated with an increased risk of ischemic heart disease, sudden cardiac death, and heart failure [5]. In general, people with AF have a four times increased risk of mortality compared to the general population [6]. The current detection of AF heavily relies on routine medical examinations; however, this approach may overlook paroxysmal AF cases, which refer to AF episodes that occur sporadically and self-terminate within 7 days. Additionally, a significant portion of AF patients, estimated at 25% to 35%, remain asymptomatic, which further reduces their likelihood of seeking care. These factors collectively contribute to delays in the identification of AF cases. Consequently, there has been a surge in efforts from both industry and academia sectors for developing technologies that enable reliable and continuous detection of AF. Such technologies are crucial for early identification and timely intervention of AF, which in turn improves patient outcomes.

To enable consistent and long-term monitoring of atrial fibrillation (AF), a solution needs to be non-intrusive, cost-effective, and convenient, reducing operational complexity and encouraging user compliance. To this end, photoplethysmography (PPG) has emerged as a preferred technology, with a ubiquitous adoption in over 71% of wearable devices given its capacity to capture heart rhythm dynamics [7]. The physiological foundation of PPG for AF detection lies in the fact that irregular heartbeats induce variations in cardiac output, leading to fluctuations in peripheral blood volume. This results in irregular pulse-to-pulse intervals and altered morphologies in PPG during AF episodes. Exploiting this physiological basis, wearables equipped with PPG sensors and specialized software offer great promise for personalized self-monitoring of AF, enabling individuals to receive timely alerts for potential AF episodes. However, the success of this approach hinges on the accuracy of PPG AF detection algorithms. Suboptimal algorithms can easily lead to a surge in false positives, thereby straining healthcare resources through unnecessary or inappropriate medical consultations.

Therefore, it marks tremendous importance for the development of precise and sensitive PPG-based algorithms for AF detection. These algorithms should aim to minimize false detections and optimize the utilization of healthcare resources, ensuring that appropriate clinical guidance is provided to individuals experiencing actual AF episodes. A prior review conducted by Pereira et al. provided a comprehensive summary of research on PPG-based AF detection using statistical analysis, machine learning (ML) and deep learning (DL) approaches up until July 2019 [8]. The review concluded that PPG holds promise as a viable alternative to ECG for AF detection. However, it also highlighted challenges such as the presence of arrhythmias other than AF, motion artifacts in PPG signals from wearable devices, and labor-intensive data annotation processes, among others.

Given the rapid technological advancements in wearable technology and methodological development in artificial intelligence (AI), there is a well-justified need for an updated review of AF detection using PPG. Building upon the previous work by Pereira et al., this paper aims to fill the gap by providing a comprehensive review of the latest developments in utilizing PPG-based digital health and AI solutions for AF detection in both inpatient and outpatient settings from July 2019 to December 2022. The articles included in this review are classified by the three methodological categories established by [8], namely, STAT, ML, and DL, to facilitate the tracking of evolving trends in the field. In addition to conducting a thorough analysis of studies on PPG-based AF detection, this study has established an online knowledge database [9]. This database encompasses all studies reviewed up to December 2022, including those from our work and Pereira's, along with direct links to the respective papers. Committed to keeping the database current, our team will update it semi-annually. Through the creation of this resource, we aim to foster community collaboration and accelerate the development of effective solutions to this critical clinical challenge.

## 2. Search Criteria

The research team used the SCOPUS, IEEE Xplore, PubMed, Web of Science, and Google Scholar databases to gather appropriate documents for the review. All articles selected were published between July 2019 and up to December 2022, and reviews were eschewed in favor of data-based research studies. Databases function similarly, but not uniformly, so queries needed to be adjusted to reflect this. Filters were used in all databases to restrict the date of publication. Table 1 describes the exact search strings used in different databases for initial document screening. After the documents were retrieved, they were further evaluated for appropriateness for review by two researchers (RX and CD). Based on this search criteria, there are in total 59 studies included in the review, including 17 STAT, 18 ML, and 24 DL studies.

| Scientific Database | Search Strings |
| --- | --- |
| **SCOPUS** | (PPG or Photoplethysmography) and (atrial fibrillation or AF or AFib or arrhythmia or cardiac rhythm) and (detection or recognition) |
| **IEEE Xplore** | ("All Metadata":atrial fibrillation) AND ("All Metadata":wearable computer) AND ("All Metadata:Photoplethysmography OR "All Metadata":PPG) |
| **PubMed** | (PPG "OR" Photoplethysmography) "AND" (atrial fibrillation "OR" AF "OR" Afib "OR" arrythmia of cardiac rhythm) "AND" (detection "OR" recognition) |
| **Web of Science** | (PPG or Photoplethysmography)(All Fields) and (atrial fibrillation or AF or afib or arrythmia or cardiac rhythm)(All Fields) and (detection or recognition)(All Fields) |
| **Google Scholar** | (PPG or Photoplethysmography) and (atrial fibrillation or AF or AFib or arrhythmia or cardiac rhythm) and (detection or recognition) |

Table 1. Search strings used in different scientific databases for study screening.

## 3. Publication trends in the past decade

Fig. 1 depicts the trends in the number of publications in the three method categories in the past 10 years between January 2013 and December 2022. It reveals an accelerated rate of growth in the number of publications in all three categories, indicating the increasing effort outpouring to developing PPG-based AF detection algorithms. It is worth noting that studies utilizing DL for AF detection emerged in 2017 and expanded rapidly, outpacing the other two categories. In the year 2022, the cumulative number of publications using DL for AF detection exceeded any of the other two categories for the first time in history.

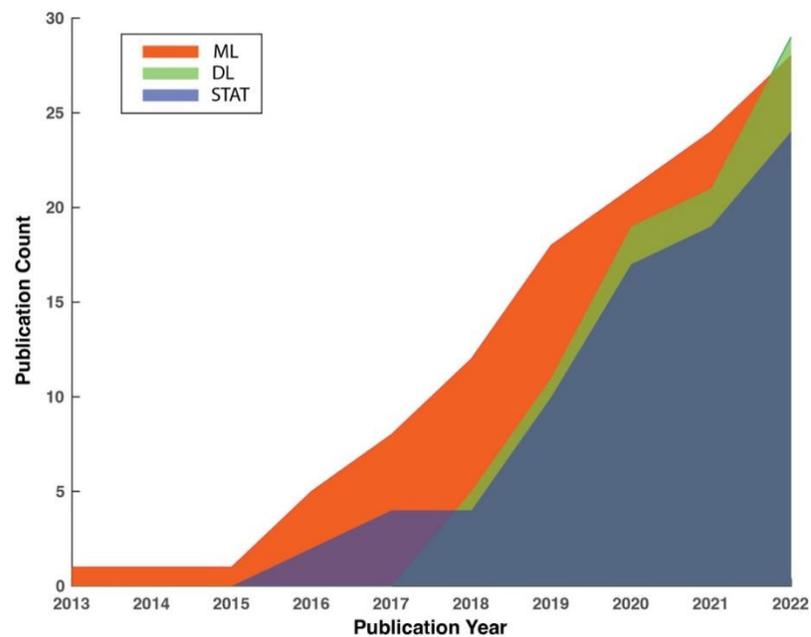

Figure 1. Trends in the number of publications in three method categories using PPG for AF detection.

Table 2. Studies on photoplethysmography based AF detection using statistical approaches

| Author (year) [ref.] | Number of patients | Dataset features | Age of population | Length PPG segments | Measurement device | Acquisition conditions | Input data | Methodology | Performance results for rhythms detection |
|---|---|---|---|---|---|---|---|---|---|
| Väliaho et al. (2019) | 213 | 106 AF, 107 NSR | 72.0 ± 14.3 YO | 5 min | Wrist band | Outpatient-checkpoint | Pulse-to-pulse interval | Two AF detection algorithms: AF Evidence and COSEn | Sen=0.962; Spe=0.981 |
| Eerikäinen et al. (2019) | 32 | 13 continuous AF, 10 non-AF | 70 ± 9 YO for continuous AF group, 67 ± 13 YO for | 30 sec | Data logger worn on the non- | Outpatient-continuous measure | Inter-pulse interval features: the percentage of inter-val differences of successive intervals greater than 70 ms (pNN70), Shannon Entropy | Logistic regression | 5-min data: Sen=0.989; Spe=0.990; Acc=0.990; 24-hour data: Sen=0.970; Spe = 0.920; Acc = 93.91% |
| Kabutoya et al. (2019) | 59 | 29 AF, 30 NSR | 66.5 ± 12.2 YO for AF group, 67.7 ± 8.0 YO for NSR group | 25 sec | Wrist-type monitor | Outpatient-checkpoint | 3 measurements for the left and right wrist based on irregular pulse peak (IPP) and irregular heartbeat (IHB) | Detection rule: "monitor AF in irregular pulse peak (IPP) 15/20/25" defined as follows: (a) IPP: \|interval of pulse wave − the average of the interval of the pulse wave\| ≥ the average of the interval of the pulse wave × 15/20/25%; (b) irregular heartbeat (IHB): beats of (IPP ≥ total pulse × 20%; and (c) the " monitor AF (IPP 15/20/25)" : ≥ 4 IHBs | Patient-level performance by IPP 15%: Sen=0.970; Spe=1; PPV=1; NPV=0.970 |
| Bashar et al. (2019) | UMass database 37, Chon Lab database 9 | UMass database: 10 AF and 27 non-AF; Chon Lab database: 9 healthy | -- | 30 sec | Wristband | Outpatient-checkpoint | Root mean square of successive differences (RMSSD) and sample entropy (SampEn) from the | Weighted average of two features and threshold-based rule | Sen=0.982, Spe=0.974 Acc=0.975 |
| Bashar et al. (2019) | 20 | 8 AF, 12 non-AF | -- | 30 sec | Wrist watch | Outpatient-checkpoint | Root mean square of successive differences (RMSSD) and sample entropy (SampEn) from the pulse intervals | Weighted average of two features and threshold-based rule | Sen=0.962 Spe=0.974 Acc=0.971 |
| Han et al. (2019) | 16 | Patients: 11 NSR and 3 with PAC/PVC, 2 with basal heart rate AF | 63-88 YO | 30 sec | Smart watch | Outpatient-checkpoint | Not an Afib detetion study but the HR estimation study using ppg | -- | -- |
| Sološenko et al. (2019) | 34 | Clinical testing database with 15 AF and 19 non-AF, plus two simulated developmental and testing databases | Clinical testing database 72.9 ± 8.9 YO for AF group, 67.5 ± 10 YO for non-AF group | 30 sec | PPG simulator | Simulation | PP or RR interval | Threshold based detector using Heaviside step function to calculate sample-entropy like index | Two sets of performance based on different signal quality thresholding of the dataset. sensitivity of 72.00% and a specificity of 99.70% when 89.20% of the database ; 97.20% and a specificity of 99.60% were achieved when increasing the |
| Han et al. (2020) | 37 | All patients have cardiac arrhythmia | 50-91 YO | 30 sec | Wrist watch | Outpatient-continuous measure | This is for PAC/PVC detector for AF patients or NSR subject, not for detecting AF | -- | -- |
| Inui et al. (2020) | 40 | Patients scheduled for cardiac surgery | Mean (SD) 70.9 YO (11.1) | 1min | Smart watch and wrist band | Outpatient-continuous measurement | This is for using ppg for pulse rate estimation in AF as compared to ECG | -- | -- |

Table 2. Studies on photoplethysmography based AF detection using statistical approaches

| Author (year) [ref.] | Number of patients | Dataset features | Age of population | Length PPG segments | Measurement device | Acquisition conditions | Input data | Methodology | Performance results for rhythms detection |
|---|---|---|---|---|---|---|---|---|---|
| Estrella-Gallego et al. (2020) | 9 | 4 AF, 9 Non-AF | 35-80 YO | 30 sec | Smartphone | Outpatient–continuous measurement | PPG signals with Offset removed and EWMA filter applied for smoothening | -- | -- |
| Väliaho et al. (2021) | 359 | 169 AF, 190 NSR | 72.2 ± 14.3 YO for AF | 1 min | Wrist band | Inpatient–checkpoint | The five pulse inverval-based variables were: mean PIN, root-mean-square values of successive | Linear logistic regression | Sen=0.964 Spe=0.963 AUC=0.993 |
| Avram et al. (2021) | 204 | 32 no history of AF, 159 paroxysmal AF, 16 with persistent AF | 62.61±11.6 YO | 5min | Smartwatch | Outpatient–continuous measurement | IBI features: the dispersion of the Poincaré plot, the standard deviation and the modified Shannon entropy | Logistic regression model | Sen=0.878 (95% confidence interval [CI] 0.836–0.910) Spe=0.974 (95% CI 97.10%–97.70%) |
| Chorin et al. (2021) | 18 | 6 AF, DM 4 HTN 8, Brugada syndrome 3, DFT after ICD implant 5 | 59.4 ± 21.3 YO | 1 min | Cardiac Sense Smartwatch | Outpatient–continuous measurement | RR and GG intervals of PPG and ECG | Threshold based defibrillation | |
| Chang et al. (2022) | 200 | 112 AF, 88 non-AF | 66.1 ± 12.6 YO | 5 min | Garmin smartwatch | Outpatient–continuous measurement | Standard deviation of normal-to-normal intervals and root mean square of successive RR interval | An undisclosed heart rate classifier | Performance based on 5-min segments: Sen=0.971, Spe=0.868 PPV of AF detection=0.897 |
| Han et al. (2022) | 35 | 23 NSR, 5 PAC/PVC, 5 Basal AF, 5 AF with RVR | 50 to 91 YO | 30 sec | Smartwatch | Outpatient–continuous measurement | Root mean square of successive differences (RMSSD) and sample entropy (SampEn) from the pulse intervals | Weighted average of two features and threshold-based rule | Not reported. AF detection is a part of the procedure for estimating HR. |
| Väliaho et al. (2022) | 173 | 76 AF, 97 NSR | AF: 77.1±9.7 YO, NSR: 67.3±15. | 1min | Wrist band | Outpatient–continuous measurement | The five-pulse interval-based variables were: mean PIN, root-mean-square values of successive differences (RMSSD), AF Evidence (AFE), Coefficient of | Linear logistic regression | 30-min time-frame performance: Sen=0.947, Spe=0.969 F1=0.954 |
| Nonoguchi et al. (2022) | 286 | 163 with high AF risk, 123 with known AF | 66 ± 12 YO for high-risk group, 67 ± 12 YO | 30 min | Wristwatch-type continuous pulse wave monitor | Outpatient–continuous measurement | Features based on pulse period (PP) values: CV, degree of variation and KS, Kolmogorov-Smirnov difference. | A rule-based algorithm using CV and KS | Patient-level performance: Sen=0.980 Spe=0.906 PPV=0.694 NPV=0.995. Interval level performance: 0.869, 0.988, 0.896, and 0.985 |

Table 3  Studies on photoplethysmography based AF detection using ML approaches

| Author (year) [ref.] | Number of patients | Dataset features | Age of population | Length PPG segments | Measurement device | Acquisition conditions | Input data | Methodology | Performance results for rhythms detection |
|---|---|---|---|---|---|---|---|---|---|
| Yang et al. (2019) | 11 | Patients referred to hospital in AF state | 63 ± 12 YO | 5, 10, 15, 20 sec | Customized wrist-type device | Inpatient | Statistical measures of Wavelet transform coeffients (mean, median, standard deviation, variance, Shannon entropy, energy, contrast, Inverse different moment, homogeneity) | Support Vector Machines with polynomial and radial-basis function kernels | Sen=0.701; Spe=0.886; Acc=0.804 |
| Neha et al. (2019) | 15 | 13 PPG records for training and 2 PPG sample for testing (MIMIC II) | – | 24 sec | Finger pulse from bedside monitors | Inpatient | Time series features: crest to crest intervals, trough to trough intervals; heart rate. | Artificial neural network (ANN), support vector machine (SVM), Logistic regression, decision trees and | Sen=0.980; Acc=0.977 |
| Fallet et al. (2019) | 17 | All patients referred for catheter ablation of cardiac arrhythmia [415 epochs of ventricular arrhythmia (VA), 1370 samples of AF and 381 | 57 ± 13 YO | 10 sec | Wrist-type device | Inpatient–continuous measurement | PPG-wave features: Adaptive organization index (AOI); Variance of the slope of the phase difference (VSPD);Permutation entropy (PE);Spectral entropy (SE);Fractional spectral radius (FSR)Spectral purity index (SPI); RR time series features: mean, standard deviation, median, Interquartile range, Minimum, | Bagging decision trees | AF vs NSR: Sen=0.997; Spe=0.924; Acc=0.981; PPV=0.979; NPV=0.989; F1=0.990. AF vs (SR&VA): Sen=0.962; Spe=0.928; Acc=0.950; PPV=0.959. |
| Guo et al. (2019) | 224 | 424 suspected AF, 227 confirmed AF | 55 to 32 YO | 45 sec | Wrist-type device | – | Peak-to-peak intervals of ppg for uniform SR, the variance, entropy derived from the peak-to-peak intervals were fluctuating for AF episodes | Threshold based ANN | Sen=0.93; Spe=0.84, PPV=0.85 |
| Zhang et al. (2019) | 375 | 20 AF, 140 NSR, Hypertension 47, diabetes 23, artery disease 14, current smoking 24 and | Mean age 53 YO | 45 sec | Wrist-type device | Inpatient–continuous measurement | Peak to Peak intervals of PPG, Kolmogorov-Smirnov test for normality of continuous variables, Normal distributions presenseted as Mean (SD), Mann- Whitney Test values for categorical values. | Boosting Algorithm | Sen=0.955; Spe=0.991; PPV=0.931; NPV=1; Kappa=0.960. |
| Buś et al. (2020) | 32 | 8 NSR recordings (total length of 240 min), 24 AF recordings (total length of 120 min); 253 AF samples; 381 NSR samples | – | 32 consecutive interbeat interval (IBI) | Finger pulse wave acquisition system Portapres 2 (FNS, Holland) | – | Mean IBI; standard deviation of IBI; SDSD (standard deviation of the successive differences between IBI);pSD50 (percentage of successive differences between IBI greater than 50 ms) | K Nearest Neighbors (KNN); Support Vector Machine with linear kernel (Linear SVM); Support Vector Machine with radial basis function kernel (RBF SVM); Decision Tree (DT) | Best performance: RBF-SVM. Sen, Spe and Acc= over 0.975 (specific performance unavailable due to graphic presentation); F1=0.985 |
| Corino et al. (2020) | 200 simulated PPG signals | 100 AF, 100 NSR | – | 20, 30, 40, 50, 100, 150, 200, 250 and 300 | PPG simulator based on phenomenological model | – | Variability analysis of IBI time series: mean, standard deviation, root of the mean squared differences of successive intervals, (rMSSD), the percentage of interval differences of successive intervals greater than x rms (pNNx, with x = [20, 50, 70]); Irregularity of IBI; | Linear SVM | Signal length (20–300 beats): Sen=0.881–0.991; Spe=0.940–1; Acc=0.913–0.995. |
| Eerikäinen et al. (2020) | 40 | 276 hours of AF, 116 hours of atrial flutter (AFL), and 472 hours of other rhythms (NSR, and sinus rhythm | Mean age in training set: 66 YO in AF, 63 YO in AFL and 69 YO in Other; | 30 sec | Wrist-type data logging device equipped with the Philips Cardio and Motion Monitoring | Outpatient–continuous measurement | IBI features: Shannon Entropy (ShEn); Normalized Root Mean Square of Successive Differences r (nRMSSD); pNN40 and pNN70; Sample Entropy (sampEn); Coefficient of Sample Entropy (CoSEn); PPG waveform features: Kurtosis; Hjorth mobility; Hjorth | Random Forest | AF vs AFL vs Other: Sen=0.976/0.845/0.981; Spe=0.9820.997/0.928; Acc=0.981/0.964/0.956. |
| Mol et al. (2020) | 149 | PPG recordings are obtained during NSR; AF: 108 records; NSR: 108 | 69±9 YO | 3 30-sec segments | Smartphone, | Inpatient | Several rhythm and signal features, such as heart rate variability parameters, peak amplitude, and other signal characteristics | SVM | AF vs NSR: Sen=0.963; Spe=0.935; Acc=0.949 |
| Millán et al. (2020) | Not mentioned | 828 NSR signals and 828 AF signals from five open Physionet datasets | – | – | Finger pulse from bedside monitors | Inpatient–continuous measurement | IBI time series features: Mean, standard deviation (STD), median absolute deviation (MAD), root mean square of successive differences (RMSSD), the standard deviation of Poincaré plot perpendicular to the line-of-identity (SD1), the standard deviation of R-R intervals (SDRR), total heart rate varibility (S). Time-frequency domain features: mean absolute value of approximation coefficients (MAVcA), average energy of approximation coefficients (AEcA), standard deviation of approximation coefficients (STDcA). | XGBoost | AF vs NSR: Sen=0.984; Spe=0.995; Acc=0.990 |

Table 3. Studies on photoplethysmography based AF detection using ML approaches.

| Author (year) [ref.] | Number | Dataset features | Age of population | Length PPG segments | Measurement device | Acquisition conditions | Input data | Methodology | Performance results for rhythms detection |
|---|---|---|---|---|---|---|---|---|---|
| Aydemir et al. (2020) | 7 | subject's signals acquired in squat, stepper and resting phase | 20 to 52 YO | 3-sec window-length PPG segments for | Wrist bracelet | -- | Mean, standard deviation, autoregressive model parameter, values of the real part and standard deviation, values of the imaginary part | K-nearest Neighbor, Naïve Bayes, and Decision Tree | Acc=0.930, CA rate= 0.890 |
| Guo et al. (2021) | 604 | Individuals at high risk for AF. | More than 18 YO | 48 sec | Huawei smart device and Holter ECG | Outpatient-continuous measurement | 1) Heart rate: minimum value of all RR intervals (MinHR), mean value of all RR intervals (MeanHR), median value of all RR intervals (MedianHR), and skewness of all RR intervals (SkRR); 2) heart rate variability: coefficient of variation of all RR intervals (CVRR); standard | XGBoost | AF vs NSR: Sen=0.821; Spe=0.974; Acc=0.935; PPV=0.914; F1=0.865; AUC=0.971 |
| Xie et al. (2021) | 21 | Healthy participants | -- | 10 sec | Wearables on forearm | Outpatient-checkpoint | Wavelet transform based features | SVM | AF vs NSR: Acc=0.983 |
| Neha et al. (2022) | 53 | Prospective Data: 15 PVC, 15 sinus tachycardia, 15 NSR, and 8 AFL Retrospective Data: 7 premature ventricular contractions. 6 sinus | 59.93 YO for PVC, 44.13 YO for sinus tachycardia, 59.93 YO for NSR, 54.87 YO for AFL | 16 sec devided to 2 segments of 8 sec | Polar OH1 optical sensor | Inpatient | | | |
| Zhu et al. (2022) | 70 | 28 patients undergo elective DC cardioversion of AF to NSR, 29 patients exhibit various arrhythmias including AF, and 13 participants with no prior | 69.3±10.9 YO for the 28 patients, and 35.4 ± 5.7 YO for the 13 participants | 5min | Samsung smartwatch | Outpatient-continuous measurement | The Poincare Plot dispersion, the modified Shannon entropy, the sample standard deviation, the sample mean, the mean stepping increment, the kurtosis, and the skew. | Hybrid Decision model with logistic regression and a heuristic statistical decision model based on Shannon | AF vs NSR: Sen=0.878; Spe=0.974. |
| Hiraoka et al. (2022) | 80 | Patients scheduled for cardiovascular surgery | Mean (SD) 65.8 YO (13.4) after excluding one | 10 min | Apple watch | Inpatient-continuous measurement | Median value of the mean and SD of PPG pulse rate | Gradient Boosting Decision Tree | AF vs Other: Sen=0.909; Spe=0.838 |
| Liao et al. (2022) | 116 | 76 patients with paroxysmal AF, 40 patients with persistent AF | 59.6 ± 11.4 YO | 10, 25, 40, and 80 heartbeats | Wrist-worn smartwatch | Outpatient-continuous measurement | PPI SD, RMSSD, Shannon entropy (SE10, SE100, and SE1000), rolling SD3, RMSSD3, and MaxFFTSD3 for AF discrimination | Random Forest | AF vs NSR: Sen=0.941; Spe=0.934; Acc=0.937; PPV=0.930; and NPV=0.939 |
| Jeanningros et al. (2022) | 42 | 42 patients refferred for catheter ablation | -- | 30 sec | Wrist bracelet | Outpatient-continuous measurement | IBI time series features: Shannon entropy (ShEn), normalized root mean square of successive differences (nRMSSD), percentage of differences of successive IBIs that exceed 40 or 70 ms (pNN40, pNN70), sample entropy (sampEn) and coefficients of sample entropy (CoSEn) of embedding dimensions 1 and 2, turning point ratio (TPR), minimum, maximum, mean and standard deviation (std) of IBI time series Frequency domain features: Horjth mobility and complexity, spectral entropy (specEn) and spectral purity index (SPI) pulse wave analysis (PWA) features: pulse foot, the systolic raise, the anacrotic notch (AN), the post-AN systolic peak, the dicrotic notch and the diastolic peak | Ridge regression, random forest, K-Nearest Neighbors and SVM | AF vs non-AF vs NSR: average Sen=0.734; Spe=0.879; Acc=0.840; PPV=0.645; NPV=0.841 |

Table 2. Studies on photoplethysmography based AF detection using DL approaches

| Author (year) [ref.] | Number of patients | Dataset features | Length PPG segments | Measurement device | Acquisition conditions | Methodology | Performance results for rhythms detection |
|---|---|---|---|---|---|---|---|
| Shen et al. (2019) | 29+53 | 13 with persistent AF, 2 with NSR, and 14 with changed rhythm, additional 53 healthy | 30 sec | Samsung wrist-wearable device | Outpatient–continuous measurement | 1D ResNeXt | 0.950 |
| Yousefi et al. (2019) | 30 | 15 with AF, 15 with NSR | 30 consecutive PPG pulses | Wrist-worn PPG monitor | Inpatient | Deep NN | All data: Sen=0.936 ± 0.216 Spe=0. 992 ± 0.180 AUC=0.996 After quality assessment: Sen=99.2± 1.3 |
| Zaen et al. (2019) | 105 | 84 from Long-Term AF Database from PhysioNet, 21 from Lausanne University | 30 sec | Tri-axis accelerometer | Outpatient–continuous measurement | RNN | Without outlier rejection: Acc=0.929 Sen=0.980 Spe=0.912 F1=0.875 With outlier rejection: Acc=0.986 Sen=1 Spe=0.978 F1=0.991 |
| Kwon et al. (2019) | 75 | 57 for persistent AF, 18 for long-standing persisten AF/ underwent successful | 30 sec | Pulse oximeter | Outpatient–checkpoint | 1D CNN | Sen=0.993 Spe=0.959 Acc=0.976 PPV=0.960 NPV=0.993 AUC=0.998 |
| Neha et al. (2019) | 15 | 13 samples for training, 2 samples for testing | 24 sec | PPG based sensors | Inpatient | Deep NN | Precision 0.960 Recall 0.950 Acc=0.954 F1=0.950 |
| Aschbacher et al. (2020) | 51 | All patients with persistent AF/Patients undergoing electrical cardioversion were | 10 sec | Smartwatch | Outpatient–continuous measurement | LSTM/ CNN | LSTML 0.954 Sen=0.810 Spe=0.921 DCNN Sen=0.985 Spe=0.880 AUC=0.983 |
| Torres-Soto and Ashly (2020) | 163 | 107 for cardioversion group, 41 for exercise stress test group, and 15 for ambulatory | 25 sec | Did not specify | Outpatient–continuous measurement | Autoencoder+1DCNN | Sen=0.980 Spe=0.99 F1=0.960 FPR=0.01 FNR=0.02 |
| Selder et al. (2020) | 60 | AF was identified in 6 (10%) subjects, of which 4 were previously undiagnosed | 60 sec | Wristband | Outpatient–continuous measurement | LSTM for quality assessment, Tree based classifer for AF detection | Sen=1 Spe=0.960 ACC=0.970 NPV=1 PPV=0.750 |
| Aschbacher et al. (2020) | 51 + 13 | 40 for algorithms training, 11 for algorithms testing/ 51 patients were enrolled | Roughly 30 seconds | Wrist-worn fitness tracker | Inpatient | Model1: Logistic regression Model2: LSTM Model 3: DCNN | Model 1: Sen=0.741 Spe=0.584 AUC=0.717 PPV=0.808 NPV=0.488 Model2: Sen=0.810 Spe=0.921 AUC=0.954 PPV=0.960 NPV=0.671 Model3: |
| Genzoni et al. (2020) | 37 | All patients are for catheter ablation procedures and wear an optical heart rate monitor device | 30 sec | Wrist-worn device | Outpatient–continuous measurement/ inpatient | GRU | Sen=1 Spe=0.966 Acc=0.979 |

Table 2. Studies on photoplethysmography based AF detection using DL approaches

| Author (year) [ref.] | Number of patients | Dataset features | Age of population | Length PPG segments | Measurement device | Acquisition conditions | Input data | Methodology | Performance results for rhythms detection |
|---|---|---|---|---|---|---|---|---|---|
| Chen et al. (2020) | 401 | All patients had a stable heart rhythm | >18 YO | 71 sec | Wristband | Inpatient and outpatient--checkpoint | PPG segment | SEResNet | Sen=0.950 Sep=0.990 Acc=0.976 PPV=0.986 NPV=0.970 |
| Kwon et al. (2020) | 100 | 81 for Persistent AF, 19 for long-standing persistent AF | ≥20 YO | 30 sec | Ring-type wearable device | Outpatient--checkpoint | PPG segment | 1D CNN | Sen=0.990 Spe=0.943 Acc=0.969 PPV=0.956 NPV=0.987 AUC=0.993 |
| Aschbacher et al. (2020) | 51 | All patients with persistent AF/Patients undergoing electrical cardioversion were sedated and remained supine | 63.6±11.3 YO | 10 sec | Smartwatch | Outpatient--continuous measurement | PPG segment | LSTM/ CNN | LSTML 0.954 Sen=0.810 Spe=0.921 DCNN Sen=0.985 Spe=0.880 AUC=0.983 |
| Cheng et al. (2020) | MIMIC-III waveform database: 30000 patients, IEEE | 60 sick subjects from MIMIC-III, 42 patients from IEEE TBME and 15h of PPG from synthetic dataset | TBME Respiratory Rate Benchark data set: children | 10 sec | ICU monitor and pulse oximeter | Inpatient and outpatient--continuous measurement | time-frequency chromatograph | CNN-LSTM | Sen=0.980 Spe=0.981 Acc=0.982 AUC=0.996 |
| Ramesh et al. (2021) | 37 | 10 with AF, 27 non-AF | -- | 30 sec | Simband | Outpatient--continuous measurement | Time domain features | CNN | Sen=0.946±0.02 Spe=0.952±0.07 Acc=0.951±0.03 F1=0.893±0. |
| Zhang et al. (2021) | 53 | 38 for NSR, 5 for persistent AF and 10 for paroxysmal AF | 66.3 ± 11.8 YO | 30 sec | Smartwatch | Outpatient--continuous measurement | PPG segment | multi-view convolutional neural network | Ave of Acc=0.916 Spe=0.930 Sen=0.908 |
| Das et al. (2022) | 175 | 108 with AF, 67 non-AF | -- | 25 sec | Wrist-worn wearable device | Outpatient--continuous measurement | PPG segment | Bayesian deep neural network | Without uncertainty threshold: Sen=0.722 Spe=0.720 Precision 0.627 F1=0.671 AUC=0.793 Without threshold: |
| Ding et al. (2022) | 139 | 126 for UCLA medical center, 13 for UCSF Neuro ICU | 18-95 YO for UCLA medical center, 19-91 YO | 30 sec | Pulse oximeter | Inpatient--continuous measurement | PPG segment | ResNet | Sen=0.928 Sep=0.988 Acc=0.961 PPV=0.985 NPV=0.943 |
| Sabbadin et al. (2022) | 4158 /88 | 56 from MIMIC database (13 AF), 32 from UQVSD database (2 AF) | -- | 10 sec | PPG device (did not find specified device name?) | Outpatient--continuous measurement | Root-Mean-Square (RMS) and the mean of Skewness and Kurtosis | Deep NN | F1=0.920 Precision 0.890 Recall 0.950 |

Table 2. Studies on photoplethysmography based AF detection using DL approaches

| Author (year) [ref.] | Number of patients | Dataset features | Age of population | Length PPG segm | Measurement device | Acquisition conditions | Input data | Methodology | Performance results for rhythms detection |
|---|---|---|---|---|---|---|---|---|---|
| Nguyen et al. (2022) | 40 | 18 with NSR, 15 with AF, and 7 with PAC/PVC | -- | 30 sec | PPG sensor patch measured on the wrist | Outpatient-- checkpoint | poincare plot | 2D CNN | Sen=0.968 Spe=0.989 Acc=0.981 |
| Liu et al. (2022) | 228 | Patients all have arrythmia | 52.3±11.3 YO | 10 sec | Fingertip PPG sensor | Outpatient-- continuous measurement | PPG segment | 1D CNN | AF Spe=0.934 Acc=0.944 PPV=0.890 NPV=0.940 |
| Neha et al. (2022) | 670 PPG signals /23 | 400 normal, 90 PVC, 90 tachycardia, and 90 atrial flutters | -- | 8 sec | ICU monitor | Inpatient | Dynamic time warping based features | Deep NN | Sen=0.970 Spe=0.970 Acc=0.960 F1=0.960 precision=0.960 |
| Ding et al. (2022) | 28539 patients, UCSF HER dataset, UCLA | Female AF 2304, Male AF 3473, Female cohort 13203, Male cohort 15330, NSR, PVC | 22 to 65 YO | 30 sec | Fingerprint, Wearable device | -- | PPG segment | Autoencoders + ResNet | AUC=0.960 |
| Kwon et al. (2022) | 35 | All patients underwent successful electrical cardioversion for AF | Mean 58.9 YO | 10 sec | Smart wring | Outpatient-- continuous measurement | PPG segment | Not specify | AUROC 0.995 Sen=0.987 Spe=0.978 FPR=0.02 FNR=0.01 |

## 4. Review of recent studies on PPG-based AF detection

### 4.1 Updates on PPG-based AF detection using statistical analysis approaches

A compilation of studies for PPG-based AF detection employing statistical analysis approaches is summarized in Table 2. The table provides an overview of these studies in chronological order, including patient cohorts, data characteristics, employed features and methods, care settings (inpatient versus outpatient), and the resultant performance outcomes. It shows that the statistical analysis approach mainly relies on threshold-based rules on the selected set of features for AF detection. Under this umbrella, the most frequently employed features for AF detection include the RR interval from the ECG and the inter-beat interval (IBI) from PPG [10-15]. Additionally, the root mean square of successive differences (RMSSD) and sample entropy (SampEn) are also among the most utilized features [16-20]. Consequently, the extracted features undergo analysis in terms of their histograms, both with and without the presence of AF and other cardiac rhythms. This analysis assists in determining optimal thresholds that effectively differentiate various rhythmic classes. Once these thresholds are established, they can be applied to the same features extracted from PPG signals.

Furthermore, the utilization of identical feature sets with alternative statistical approaches, such as logistic regression, enhances the versatility and comprehensiveness of AF detection studies. By applying logistic regression, researchers can establish a mathematical model that estimates the probability of AF presence based on the input features. The logistic function, also known as the sigmoid function, is employed to transform the output into a range between 0 and 1. This transformed probability serves as an indicator of the likelihood of AF compared to non-AF cases. The advantage of logistic regression lies in its ability to provide a quantitative measure of the probability, allowing for a nuanced understanding of the classification outcome. Also, as reported in Table 2, studies incorporating larger patient cohorts intend to utilize logistic regression [14, 16, 19] rather than rule-based models. This observation aligns with the trends identified in a previous review study [8], further reinforcing the preference for logistic regression in cases involving a higher number of patients.

As compared to the previous review, we observe a rising number of studies using the statistical analysis approach (4.25 studies/year between 2019 and 2022 vs. 2 studies/year between 2013~2019), which aligns with the rising number of all-type AF detection studies in recent years. It can be observed that more studies focus on outpatient populations, which might be attributed to the rapid advancement of wearable technology in recent years.

**4. 2 Updates on PPG-based AF detection using machine learning approaches**

Table 3 presents a chronological summary of AF detection studies based on machine learning approaches in the last four years. Machine learning has demonstrated promising results in the detection of AF in low-sample settings. The application of ML techniques requires domain expertise for feature engineering to extract features that effectively capture the comprehensive characteristics of PPG waveforms and enable the discrimination of different classes. Commonly extracted features include morphological descriptors, time domain statistics, statistic measurements in the frequency domain, nonlinear measures, wavelet-based measures, and cross-correlation measures.

Of different machine learning algorithms, Tree-based algorithms, such as Decision Trees, Random Forest, and Extreme Gradient Boosting (XGBoost) [21], are the most popular choices and are collectively employed in 12 out of the 18 studies employing machine learning for AF detection. Random Forests have demonstrated strong performance in AF detection tasks using PPG. This ensemble learning algorithm combines multiple decision trees to create a robust classification model. By aggregating the predictions of individual trees, Random Forests can reduce overfitting, handle complex feature interactions, and provide accurate AF detection results. The versatility, interpretability, and resilience to noisy data make Random Forests a popular choice in PPG-AF detection research. XGBoost is a boosting algorithm that combines gradient boosting with decision trees to achieve high predictive accuracy in PPG-AF detection. XGBoost sequentially builds an ensemble of weak models, iteratively improving its performance by minimizing a loss function. It can effectively handle complex feature interactions and capture subtle patterns in PPG signals, leading to improved AF classification results and better detection performance compared to individual decision trees.

The second most popular (used in 8 out of 18 studies) machine learning classifier for AF detection is Support Vector Machines (SVM) [22], due to their ability to handle high-dimensional feature spaces. SVM separates PPG signal data into different classes by identifying an optimal hyperplane that maximizes the margin between the classes. By mapping PPG signals into a higher-dimensional space, SVM can capture complex relationships and find effective decision boundaries for accurate AF classification. There are also other classifiers adopted in the studies such as K-Nearest Neighbors (KNN) and artificial neural networks (ANN) but are not widely adopted as the above two classifiers.

Compared to the previous review, we observe a sharp increase in the adoption of machine learning for AF detection using PPG (5 studies/year between 2019 and 2022 vs. 1.5 studies/year between 2016 and 2019).

## 4.3 Updates on PPG-based AF detection using deep learning approaches

Deep learning has emerged as a powerful approach for detecting AF in PPG signals, as reported in Table 4. Unlike traditional ML methods, DL models can learn comprehensive feature representations through an end-to-end learning fashion, eliminating the need for complex feature engineering. This is achieved by learning from a large amount of training samples to train deep neural networks, which consist of interconnected layers of computational nodes.

As shown in Table 4, studies using DL approaches can be divided into two main categories. The first category (employed in 14 out of 24 studies) is a family of convolutional neural networks (CNN).

CNN is commonly applied in computer vision tasks, but they have also been successfully adapted for PPG-AF detection CNNs utilize convolutional layers to automatically extract relevant features from the PPG signal data [23]. These convolutional layers apply numerous filters across the signal, allowing the network to capture local patterns and identify important discriminative features associated with AF. By stacking multiple layers, CNNs can learn increasingly complex representations of the PPG signals, enhancing the accuracy of AF detection. Residual network (ResNet) [24], a specific type of CNN, addresses the challenge of training deep neural networks by utilizing skip connections. These connections allow the network to bypass layers and pass information directly to subsequent layers, mitigating the vanishing gradient problem. In the context of PPG-AF detection, ResNet architectures enable the training of deeper networks with improved performance and ease of optimization. By incorporating residual connections, ResNet models can capture fine-grained details and long-range dependencies in PPG signals, leading to enhanced AF detection capabilities. The second category is a family of sequential DL models, of which Long Short-Term Memory ( LSTM ) [25], is a popular choice (employed in 4 out of 24 studies). LSTM is a recurrent neural network architecture commonly used in PPG-AF detection due to its ability to effectively capture temporal dependencies in sequential data. In the context of PPG signals, LSTM models can analyze the sequential nature of the data, considering the temporal order of the signal samples. This allows LSTM to capture long-term patterns and dynamic changes in the PPG signals, which are crucial for accurate AF detection.

To effectively train DL models, a substantial amount of labeled training data is typically required. However, in biomedical applications, the availability of labeled data is often limited. Transfer learning is a potential solution to this challenge, wherein a pre-trained DL model is fine-tuned for a specific task. The number of layers and the complexity of fine-tuning depend on the particular application. For example, in one study, a pre-trained CNN model designed for ECG analysis was fine-tuned to detect AF from PPG segments using a small set of labeled data. Another promising technique is data augmentation to generate artificial samples to boost the number of samples for training the DL models and increasing the generalizability of model performance.

DL is the fastest growing approach of all three approaches for PPG-AF detection. We observe an average of 6 studies employing DL per year between 2019 and 2022, as compared to 3.5 studies/year between 2018 and 2019.

## 5. Discussion

While the performance metrics reported in Tables 2-4 suggest the promising potential of PPG for AF detection, several challenges remain. In this section, we will delve into these issues, offering insights drawn from our comprehensive analysis of the reviewed studies. Key concerns to be discussed include PPG signal quality, label accuracy, and the impact of concurrent arrhythmias. Studies that have considered these issues are summarized in Table 5. Furthermore, we extend our discussions to encompass additional considerations pertaining to PPG-based AF detection. These include algorithmic factors such as performance metrics, data sources, computational efficiency, domain shifts, as well as model explainability and equity.

| Factors | Studies | | Capacity |
|---|---|---|---|
| Signal Quality | STAT | [16],[11],[17],[[18],[26],[12],[27],[28],[13],[14],[15] | 11/17 |
| | ML | [29],[30],[31],[32],[33],[34],[35],[36],[37] | 9/18 |
| | DL | [38],[39],[40],[41],[42],[43],[44],[45],[46],[47],[48],[49] | 12/24 |
| Label Noise | STAT | [10],[28],[13],[15] | 4/17 |
| | ML | [30],[35],[50],[36] | 4/18 |
| | DL | [40],[51],[43],[47],[52],[48] | 6/24 |
| Concurrent Arrhythmias | STAT | [16],[17],[26],[27],[14] | 5/17 |
| | ML | [31],[36] | 2/18 |
| | DL | [40],[53],[48],[54] | 4/24 |

**Table 5.** Challenging factors considered in the studies

### 5.1 PPG signal quality

PPG signal quality remains a considerable challenge, which is widely acknowledged within the scientific community. A multitude of complicating factors can compromise the PPG signal quality, including motion artifacts, skin tone variations, sensor pressure variations, respiratory cycles, and ambient light interference, only to name a few. The challenge of noise in PPG signal is particularly acute when it comes to the continuous acquisition of PPG, which is crucial for long-term monitoring of AF risk.

As reported in Table 5, most of the reviewed studies take signal quality into consideration, with 54% of the reviewed studies implementing measures to exclude PPG signals of poor quality. For example, in [27], the authors presented a noise artifact detection algorithm designed for detecting noise artifacts. Out of a total of 2,728 30-second PPG strips, only 314 strips were deemed suitable for further analysis after applying the algorithm. Similarly, in [41], the authors proposed a multi-tasking framework that incorporated both signal quality assessment and AF detection tasks. Only PPG signals of excellent quality were retained for the purpose of AF detection. This practice, however, harbors potential issues that warrant deeper consideration. Firstly, by systematically discarding vast swaths of signal data considered of inferior quality, the earliest possible detection of atrial fibrillation (AF) is inevitably delayed, creating a potentially significant time lag in diagnosis. Secondly, this approach harbors a statistical dilemma; the discarded PPG-AF signals could be construed as false positives within the context of the overall analysis. However, such instances are typically overlooked when calculating the positively predicted value or false positive rate, thereby potentially inflating the model's reported performance. Consequently, the reliance on selective data exclusion as a signal quality control strategy may inadvertently compromise the validity of the study's outcomes and the efficacy of predictive calculation would be based on the proportion of motion artifacts present within individual PPG segments, thus providing a more precise estimate of signal quality. Subsequently, an appropriate threshold could be ascertained to filter out PPG signals devoid of meaningful information. Alternatively, one can integrate the signal quality information as part of the model input that controls the uncertainty level of the model output. These approaches would strike the balance of salvaging PPG signals with suboptimal quality for disrupt-less monitoring and model performance.

**5.2 Label noise**

The issue of label noise in annotated datasets presents another significant challenge in the application of PPG for AF detection. Accurate and consistent labeling of datasets is crucial for the development and validation of reliable detection algorithms [56]. To achieve this, it usually involves more than two clinical domain experts to cross-check the agreement of annotations, and a reconciliation strategy needs to be in place in the event of disagreement. However, many studies often fall short in this aspect due to the labor-intensive task and an insufficient number of cardiologists available to annotate the datasets. Across the reviewed studies, only 9 out of the 59 studies [10, 13, 15, 28, 35, 36, 40, 47, 48] employed the expertise of at least two cardiologists for annotation, as reported in Table 5. This scarcity of expert annotators can result in imprecise and incomplete labeling of AF events, leading to label noise, which in turn, may undermine the performance of supervised learning algorithms.

Furthermore, the absence of standardized guidelines to address disagreements among annotators exacerbates this issue. In the event of conflicting annotations, the lack of a clear protocol or consensus mechanism can lead to inconsistencies in the dataset. This variability not only confounds the training of predictive models but also hampers the reproducibility of research findings. Consequently, establishing robust procedures for data annotation, which involve recruiting sufficient expert annotators and defining clear rules for resolving disagreements, is paramount. Addressing these issues would significantly enhance the quality of the annotated PPG datasets, thereby facilitating more reliable and accurate AF detection.

In addition to the shortage of expert involvement, the field faces another substantial challenge: the absence of clear clinical guidelines for annotating AF events using PPG data. Unlike ECG, which has well-established guidelines for AF event labeling, PPG operates in a far less standardized environment. This lack of formalized guidance further exacerbates the risk of label noise, compromising both algorithmic performance and clinical reliability. Given these constraints, it becomes imperative to consider multimodal signal inputs when annotating data. Incorporating ECG or other established modalities alongside PPG can provide a more robust framework for annotation, thereby improving the quality of labeled data.

### 5.3 Concurrent arrhythmias

The detection accuracy of AF through PPG can be significantly influenced by the presence of other arrhythmias, notably premature ventricular contractions (PVC) and premature atrial contractions (PAC). Both PVCs and PACs introduce irregularities into the heart rhythm that can mimic the rhythm irregularities seen in AF, potentially leading to false-positive detections. PVCs and PACs are characterized by early heartbeats originating from the ventricles and atria, respectively [27]. These early beats can disrupt the regular rhythm of the heart, resulting in PPG signal patterns that may resemble those associated with AF. Consequently, a PPG-based AF detection model might mistakenly classify these as AF events, thereby reducing the specificity of the model. Furthermore, the simultaneous presence of AF and other arrhythmias in the same patient adds another layer of complexity to the problem. This co-existence can modify the PPG signal's morphology in ways that differ from the signals of patients with AF or PVC/PAC alone, making it more difficult to accurately identify the presence of AF.

It is noteworthy that several studies considered the presence of arrhythmias other than AF, as shown in Table 5. For instance, in the study by [31, 36], the differentiation of PVC and PAC from AF using PPG signals was explored. The results of this investigation demonstrated successful differentiation between PVC/PAC and AF based on PPG signal characteristics. This finding suggests that PPG-based analysis holds promise for distinguishing various types of arrhythmias beyond AF. Thus, when developing and evaluating PPG-based AF detection models, it is critical to account for the potential influence of other arrhythmias like PVC and PAC. Robust algorithms should be designed to discriminate between AF and these other rhythm disturbances to maintain high detection accuracy, reinforcing the necessity of comprehensive, diverse, and well-annotated training datasets in the development of these predictive models.

### 5.4 Quantitative metrics for algorithm performance evaluation

The studies reviewed in this work always use conventional performance metrics, such as the area under the receiver operational characteristics curve (AUROC), accuracy, sensitivity, specificity, and F1 Score. However, it's crucial to acknowledge that relying solely on these conventional metrics may be insufficient, particularly within the context of continuous health monitoring scenarios [57]. The landscape of continuous health monitoring, facilitated through wearable devices, unfolds as a dynamic and perpetually evolving terrain of data. Within this context, the intrinsic nature of a continuous data stream introduces complexities that transcend the conventional boundaries of traditional evaluation metrics. In scenarios wherein health-related parameters undergo ceaseless scrutiny, the spectrum of fluctuations, subtleties, and overarching trends assumes paramount significance. Conventional metrics, by design, tend to compartmentalize performance assessment within discrete segments, potentially missing the panoramic context that is intrinsic to continuous health monitoring. This paradigm invites us to reflect upon the necessity of embracing evaluation methodologies that are attuned to the temporal dynamics, such as assessing the frequency of AF occurrence that reflects AF burden, the nuances of variation, and the holistic import of trends.

### 5.5 Domain shift problem

models developed therefrom.

We propose a nuanced perspective on PPG signal quality assessment rather than adhering to the dichotomous approach of designating signals as merely black or white [7]. Instead, we suggest the computation of a Signal Quality Index (SQI) as a continuous metric [55]. This

PPG signals, despite their utility in non-invasive physiological monitoring, present certain complexities linked to the site of acquisition and inter-patient variability. It has been observed that PPG signals sourced from distinct anatomical sites yield diverse morphological patterns [58]. This is primarily due to the different vascular structures, skin thickness, and other physiological attributes specific to these sites. Such morphological variations can pose significant challenges in interpreting these signals and developing universally applicable models, as the distribution of signal characteristics is inherently contingent on the site of collection.

Moreover, inter-patient variability further compounds this issue by introducing additional variations in the data distribution. These variations stem from a wide array of factors, including demographic attributes (such as age and sex), physiological characteristics (including skin pigmentation and body mass index [BMI]), and medical conditions unique to individual patients [59]. For instance, an older patient might exhibit a different PPG signal morphology due to increased arterial stiffness, while individuals with darker skin might present a different signal-to-noise ratio owing to higher melanin content that can observe more light than lighter skin.

These site-specific and inter-patient differences can induce what is referred to as a "domain shift" problem in machine learning [60, 61]. Here, a model that is trained on data from a specific group (for example, PPG signals from a certain body site or a particular patient group) may not generalize the model performance when it is applied to a different group. Therefore, while harnessing PPG signals for health monitoring and disease prediction, it is paramount to consider these variations and devise strategies to address the domain shift problem for reliable and generalized model performance.

### 5.6 Lack of large-scale labeled dataset

In concert with the label noise issue discussed in Section 5.2, there exists a challenge of a paucity of large-scale, annotated datasets. To develop robust and reliable algorithms for AF detection, especially when deep learning models are employed, it requires extensive, labeled datasets. These ideal datasets should encompass a broad range of patient demographic groups, diverse health conditions, and various physiological states to ensure generalizable findings. Furthermore, they should contain precise annotations of the AF events in the PPG signal to facilitate effective supervised learning.

Emerging research is increasingly focused on addressing this issue by generating synthetic PPG signals through various data augmentation techniques. These range from traditional computational models that simulate physiologic PPG patterns (e.g., PPGSynth) [62] to advanced generative models such as Generative Adversarial Networks (GANs) [63, 64], Variational Autoencoders (VAEs) [65], and diffusion models. However, the extent to which these synthesized signals contribute to improved learning outcomes remains an open question. Recent research by Cheng et al. indicates the existence of a 'performance ceiling'—a limit to the improvements achieved by incorporating synthetic signals [64]. This underscores the need for further investigation into more effective algorithms for synthetic signal generation as well as a deeper understanding of this performance ceiling phenomenon.

To sum up, the lack of large, labeled datasets impedes the progress of research in this area, limiting the development and validation of predictive models. It restricts the ability to comprehensively evaluate and compare the performance of different AF detection methods under diverse and challenging conditions. Additionally, it hampers the exploration of more advanced machine learning techniques, which often necessitate large quantities of annotated data to train effectively. Therefore, efforts to collect/generate, share, and consolidate large-scale, well-annotated PPG datasets for AF detection represent a critical step to move the performance needle in this field.

## 5.7 Computational time

With the rapid advancement of graphics processing units (GPUs) and increasing computational power, it is now feasible to train complex, large-scale neural networks that outperform traditional statistical or conventional machine learning methods [66]. However, this complexity presents new challenges, particularly for model inference. The inference process, which involves generating predictions from new data based on trained models, can be computationally demanding. This poses significant obstacles for wearable technologies that rely on edge computing, as these calculations can quickly deplete battery life, thereby undermining the feasibility of continuous monitoring [67]. Alternative solutions include offloading computational tasks to more powerful, tethered smartphones or to cloud-based platforms. Yet, both alternatives require robust and fast data streaming infrastructures.

Research efforts to address these challenges are bifurcated. On one hand, there is a burgeoning focus on 'tiny ML,' which aims to optimize neural network architectures for efficient edge computing without sacrificing performance. On the other hand, advancements in hardware and battery technology are driving the development of more powerful sensing techniques that enhance the capacity for long-term monitoring. Consequently, tackling these computational challenges necessitates orchestrated efforts from both research directions. It also underscores the imperative to keep computational requirements at the forefront when developing PPG-based AF detection algorithms.

## 5.8 Explainability

Explainability in the context of PPG AF detection algorithms is a critical aspect that determines how well we understand the decision-making process of these algorithms. This is particularly important in healthcare, where the decisions made by these algorithms can have significant implications for patient care. Statistical methods are often considered naturally explainable because they rely on well-understood mathematical principles and procedures. For example, a linear regression model, which lies in the intersection between statistical methods and machine learning, makes predictions based on a weighted sum of input features. The weights (or coefficients) assigned to each feature provide a direct measure of the feature's importance in the prediction, making it relatively straightforward to interpret the model's decisions. Machine learning methods, on the other hand, often involve more complex computations and may not be as directly interpretable as statistical methods. However, techniques have been developed to calculate feature importance, which can provide a certain level of explainability. For instance, in [68], the Fisher score method was employed to calculate the importance of features. The Fisher score is a statistical measure that evaluates the discriminative power of individual features in a classification task. By utilizing this method, the study aimed to assess the relevance and significance of different features in the context of atrial fibrillation detection. Similarly, in [37], each feature was input into the classifier separately, enabling the generation of a ranked list based on its impact on the overall classification performance through this sensitivity analysis.

Deep learning models, on the other hand, are often referred to as 'black boxes,' which make predictions based on intricate, high-dimensional mappings that are difficult to comprehend for humans. While they may achieve high predictive accuracy, it's often challenging to understand what features and their interactions the models use to make predictions, and how these features contribute to the final decision. This lack of transparency can be a major drawback in healthcare applications, where it's desirable to understand the underlying decision logic so as to gain trust from end users, such as clinicians and patients.

Several approaches are being explored to improve the explainability of deep learning models, including attention mechanisms, layer-wise relevance propagation, and model-agnostic methods like LIME (Local Interpretable Model-Agnostic Explanations) and SHAP (SHapley Additive exPlanations) [69-72]. A good example is [48], where authors used the guided gradient-weighted class activation mapping (Grad-CAM) approach to visualize crucial regions within the PPG signals that enabled the model to predict a specific rhythm category. Despite these advances, explainability in deep learning remains an active area of research, particularly in the context of PPG-based AF detection.

## 5.9 Performance Bias and Model Equity

Disparities in both access to and outcomes from utilizing digital health solutions and biotechnologies manifest a variety of identity dimensions, including economic status, social background, ethnicity, and gender [73]. As described by [74], health equity means, "...striving for the highest possible standard of health for all people and giving special attention to the needs of those at greatest risk of poor health, based on social conditions." In the context of PPG-based AF detection, this issue of equity extends across a spectrum of potential causes. It encompasses accessibility issues, particularly for individuals from rural areas or those with disadvantaged socioeconomic statuses, as well as physiological factors like skin tone and obesity, which can influence the reliability of PPG readings [75, 76]. Of the studies reviewed, a mere three explicitly touched upon the issue of performance bias and model equity [19, 45, 51]. This oversight underscores the pressing need to heighten awareness and equity considerations within the field. To tackle this challenge, a multidisciplinary approach is necessary, and healthcare providers, engineers, and researchers must proactively develop technologies that consider the needs of vulnerable and underrepresented populations.

## 6. Conclusion

In conclusion, this comprehensive review highlights the growing significance of PPG-based AF detection in addressing a critical clinical challenge. The surge in research efforts, especially in machine learning and deep learning approaches, underscores the potential of PPG technology for continuous and accurate AF monitoring. While machine learning techniques offer versatility and promising results, deep learning models demonstrate remarkable performance by automating feature extraction. Nevertheless, challenges related to signal quality, label accuracy, and concurrent arrhythmias persist, necessitating ongoing research and development. Furthermore, the availability of large-scale labeled datasets, computational efficiency, model explainability, and addressing performance bias and equity issues emerge as crucial considerations in advancing PPG-based AF detection technology. This review underscores the importance of continued collaboration between the medical and artificial intelligence communities to refine and deploy effective solutions for AF detection, ultimately improving patient outcomes in the face of this widespread health concern.

**Acknowledgment**

This work was partially supported by NIH grant award R01HL166233.